\documentclass[pdflatex,sn-mathphys-num]{sn-jnl}

\usepackage{graphicx}%
\usepackage{multirow}%
\usepackage{amsmath,amssymb,amsfonts}%
\usepackage{amsthm}%
\usepackage{mathrsfs}%
\usepackage[title]{appendix}%
\usepackage{xcolor}%
\usepackage{textcomp}%
\usepackage{manyfoot}%
\usepackage{booktabs}%
\usepackage{algorithm}%
\usepackage{algorithmicx}%
\usepackage{algpseudocode}%
\usepackage{listings}%
\usepackage{subfigure}
\usepackage{booktabs}
\usepackage{array}

\theoremstyle{thmstyleone}%
\theoremstyle{thmstyletwo}%

\theoremstyle{thmstylethree}%

\begin{document}

\title[Article Title]{Learning-based safety lifting monitoring system for cranes on construction sites}


\author[1]{\fnm{Hao} \sur{Chen}}\email{hao.chen@ust.hk}

\author[1]{\fnm{Yu Hin} \sur{Ng}}\email{sampsonng@ust.hk}

\author[1]{\fnm{Ching-Wei} \sur{Chang}}\email{ccw@ust.hk}

\author[1]{\fnm{Haobo} \sur{Liang}}\email{hbliang@ust.hk}

\author*[1]{\fnm{Yanke} \sur{Wang}}\email{yankee.wann@gmail.com}\email{yankewang@ust.hk}

\affil[1]{\orgdiv{Hong Kong Center for Construction Robotics}, \orgname{The Hong Kong University of Science and Technology}, \orgaddress{\street{Units 808 to 813 and 815, 8/F, Building 17W, Hong Kong Science Park}, \city{Pak Shek Kok}, \state{New Territories}, \country{Hong Kong SAR}}}





\abstract{Lifting on construction sites, as a frequent operation, works still with safety risks, especially for modular integrated construction (MiC) lifting due to its large weight and size, probably leading to accidents, causing damage to the modules, or more critically, posing safety hazards to on-site workers. Aiming to reduce the safety risks in lifting scenarios, we design an automated safe lifting monitoring algorithm pipeline based on learning-based methods, and deploy it on construction sites. This work is potentially to increase the safety and efficiency of MiC lifting process via automation technologies. A dataset is created consisting of 1007 image-point cloud pairs (37 MiC liftings). Advanced object detection models are trained for automated two-dimensional (2D) detection of MiCs and humans. Fusing the 2D detection results with the point cloud information allows accurate determination of the three-dimensional (3D) positions of MiCs and humans. The system is designed to automatically trigger alarms that notify individuals in the MiC lifting danger zone, while providing the crane operator with real-time lifting information and early warnings. The monitoring process minimizes the human intervention and no or less signal men are required on real sites assisted by our system. A quantitative analysis is conducted to evaluate the effectiveness of the algorithmic pipeline. The pipeline shows promising results in MiC and human perception with the mean distance error of 1.5640 m and 0.7824 m respectively. Furthermore, the developed system successfully executes safety risk monitoring and alarm functionalities during the MiC lifting process with limited manual work on real construction sites.}




\maketitle

\section{Introduction}
\label{sec:introduction}
Tower cranes, as one of the core equipment in the modern construction industry, play an indispensable role in high-rise buildings and large-scale projects. With their powerful lifting capacity and extensive operational range, tower cranes enable efficient and precise transportation of construction materials to designated locations, significantly enhancing construction efficiency and reducing labor costs. Additionally, their modular design and flexibility allow them to adapt to various complex construction environments, ensuring safety and progress on construction sites. Whether for high-rise buildings, bridge construction, or large infrastructure projects, tower cranes are essential for the smooth execution of engineering tasks. In addition, Modular Integrated Construction (MiC) significantly increases the demand for tower cranes due to its unique construction approach~\cite{wuni2020critical}. In MiC, prefabricated modules are manufactured off-site and then transported to the construction site for assembly. Tower cranes are essential for lifting and precisely positioning these large, heavy modules, ensuring efficient and safe installation. The high lifting capacity and extensive reach of tower cranes make them ideal for handling the size and weight of MiC modules, while their flexibility allows them to operate in constrained urban environments. As MiC gains popularity for its speed, quality, and sustainability, the reliance on tower cranes becomes even more critical, highlighting their vital role in the future of modern construction.


However, even though tower cranes are so widely used, the utilization of them also introduces a unique set of risks that must be carefully managed to ensure the worker safety and the overall success of construction projects. One of the concerns is the risk of falling objects, which can arise from improper rigging, inadequate load handling, or mechanical failures~\cite{jiang2020safety}. Such incidents can lead to the uncontrolled dropping of materials, which poses a significant danger to workers and bystanders below. Similarly, for MiC, even though this method offers several potential benefits, such as increased efficiency, improved quality control, and reduced on-site disruptions, making it an attractive option for many construction projects~\cite{tsz2023critical}, it also introduces a unique set of risks that must be carefully considered and managed~\cite{khan2021systematic}. Due to the considerable weight and size, these components require specialized lifting equipment and skilled operators to ensure safe transportation and installation. Inadequate lifting procedures can lead to accidents, damaging the modules and posing safety hazards to workers on site~\cite{zhang2022critical}. Monitoring MiC lifting is more necessary due to its weight and size. 


\begin{figure*}[!htbp]
\centering
\includegraphics[width=13 cm]{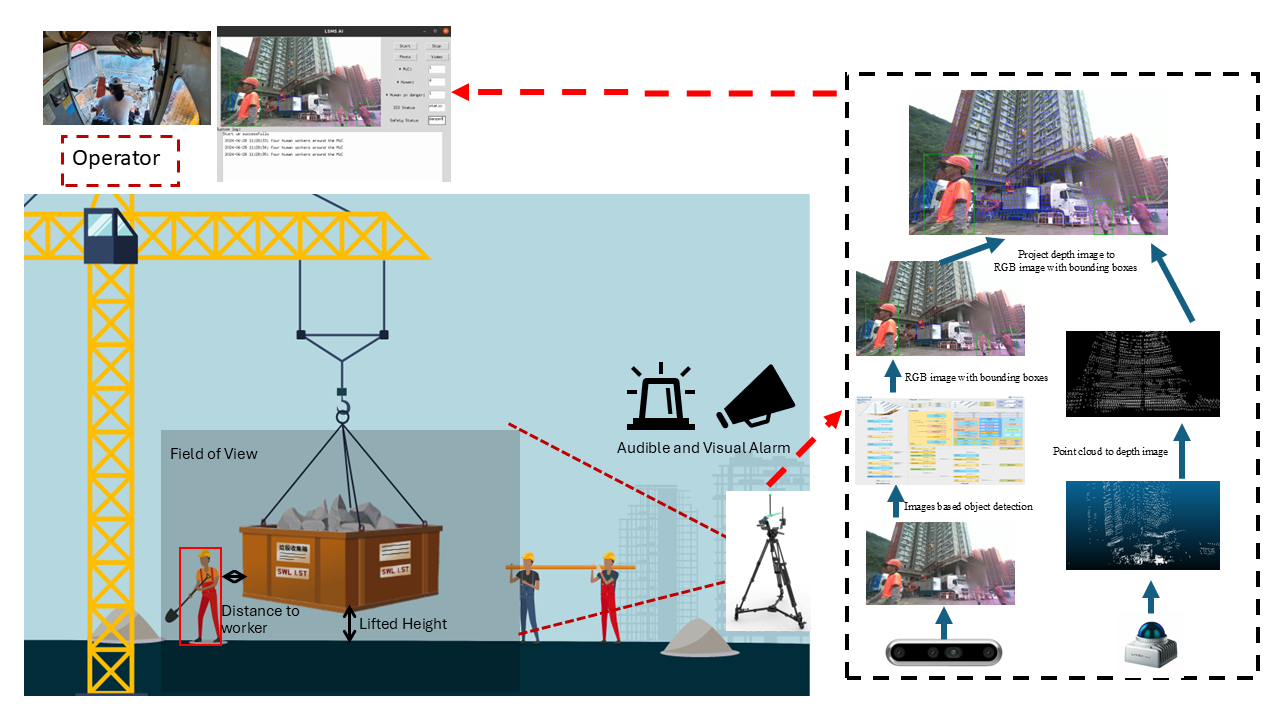}
\caption{ The usage scenario and pipeline of developed safe lifting monitoring system. The usage scenario of such system is for monitoring crane's MiC lifting. Because the position used for MiC lifting is generally fixed, the system can be placed next to the MiC via a tripod. Such system is equipped with a camera and LiDAR for perceiving that if there is human in the MiC lifting danger zone. This system can issue an audible and visual alarm if the safety hazard is detected. Meanwhile,  This video stream and safety signal will be sent to the crane operator to eliminate blind spots and give early warning.}
\label{fig:usage scenario}
\end{figure*}

In recent years, the government has taken proactive measures to improve construction site safety. In Hong Kong, the Development Bureau (DEVB) has been committed to driving a wider adoption of the smart site safety system (4S) in the construction industry to provide a safe working environment for site personnel. The Hong Kong Construction Industry Council also proposes a safety lifting procedure "3-3-3", which can effectively improve the safety of lifting operations. Such procedures include: Keep 3 meters away from materials being lifted; Lift up the materials 0.3 meters from ground; and wait for 3 seconds for stabilizing the lifting object before lifting operation. Similarly, there are also a series of rules about safe lifting proposed by Occupational Safety and Health Administration in the United States. Those policies about safe lifting for tower cranes show that there are still many safety risks in lifting. In addition, research in \cite{kim2020importance} shows that the operator’s visibility of the location of target object and the surrounding conditions are more important than any other tower crane accident factors. Thus, it is necessary to develop a system for the tower cranes' safe lifting, which can eliminate the visual blind area of tower crane operators and detect potential safety hazards. With the growing popularity of MiC lifting, the demand for such safety lifting monitoring systems will be further enhanced. 

The safety lifting path generation is a vital topic for the tower crane safe lifting and substantial efforts have been made over the years in this regarding~\cite{dutta2020automatic,hu2021automation,zhu2022crane}. However, the focus of this article is on the relative position relationship between personnel and MiC and detect potential risks of lifting initial phase, which is usually ignored by research on safe lifting path generation. There are many studies about the safe lifting monitoring system of tower cranes and the details are shown in the next section. However, the existing research has some limitations: 1) Much of the safety monitoring with AI capabilities is based on two-dimensional images and lacks three-dimensional information, 2) The existing safe lifting monitoring systems lack identification and judgment of risks, 3) There are few safety monitoring systems specially developed for cranes with MiC. For instance, \cite{zhou2019cyber} employed an IoT-based safety monitoring system for blind hoisting, and \cite{liangdigital} introduced a digital twin method based on 4D point cloud sensing and as-designed BIM to effectively improve the safety and efficiency of the MiC hoisting process.

As shown in Fig.~\ref{fig:usage scenario}, we develop a learning-based safe monitoring system for tower cranes, especially for MiC lifting, on the construction site. For implementing such a system, there are many challenges: 1) Collecting data on construction sites is difficult due to safety concerns; 2) The complex and variable site scenes pose challenges to the accuracy and robustness of the object detection algorithms; 3) Safety hazards on construction sites are numerous and diverse, making it challenging to detect as many as possible. The contribution of this paper is as follows: 1) We design an automated three-dimensional (3D) learning-based safe monitoring pipeline using the camera and LiDAR for modular integrated construction (MiC) lifting process.; 2) A dataset is created which consists of 1007 image-point cloud pairs about 37 MiC liftings along with its annotations; 3) An automated software framework based on object detection is implemented and applied to our hardware system; 4) We verify effectiveness of the proposed system on the real construction sites, proving it can automatically work to ensure a safe MiC lifting process; 5) Such system reduce the human (signal men) intervention for the MiC lifting procedures with automatic warning trigger, minimizing the possible safety hazards to workers on site.

The reminder of this paper is as follows: Section~\ref{sec:related_work} gives the related work about safety monitoring on the construction sites. Section~\ref{sec:method}  introduces key methods of the proposed system algorithm pipeline in this paper. The data processing is described in Section~\ref{sec:data_processing}. The experimental results are given in Section~\ref{sec:experiment}. Conclusion and future work are shown in Section~\ref{sec:conclusion}.

\begin{figure*}[!htp]
\center
\includegraphics[width=13 cm]{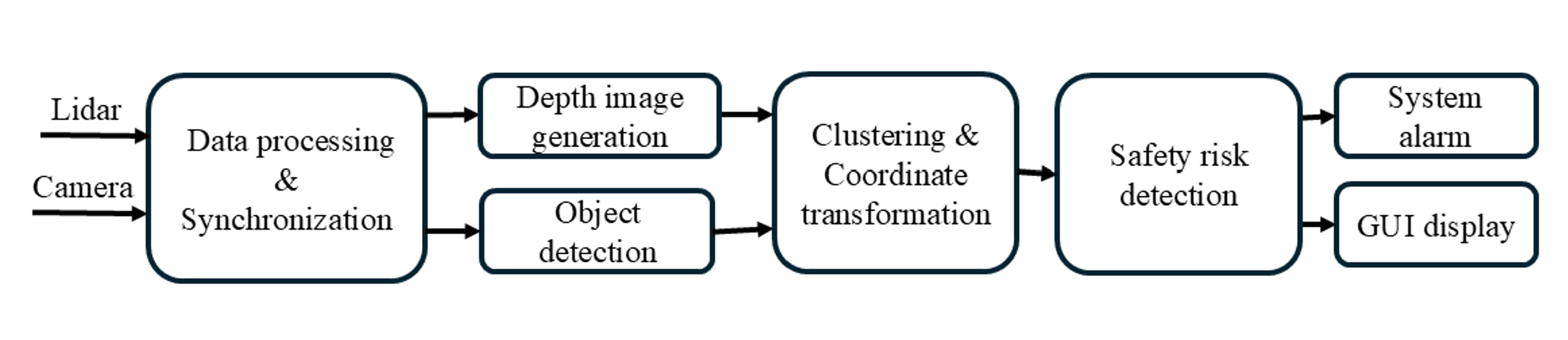}
\caption{The algorithm pipeline of developed safe lifting monitoring system. Initially, the system begins with data processing and synchronization. Then processed and synchronized point clouds are utilized to generate depth images, which provide critical spatial information about the surroundings. Meanwhile, the system performs object detection to identify MiC and humans within the scene. The processed data undergoes clustering and coordinate transformation, allowing the system to find detected objects' depths and translate their positions into a unified coordinate system. The audible and visual alarm will be triggered if safety risk detection block detects humans in the MiC lifting danger zone. while simultaneously updating the GUI display to provide operators with real-time visual feedback.}
\label{fig:pipeline}
\end{figure*}

\section{Related Work}\label{sec:related_work}
There are wide research and effort aiming to improve construction site safety. For example, \cite{yang2020automated} develops an automated personal protective equipment (PPE) and tool pair checking system using the internet of things (IoT) with wireless WiFi modules tagged on the PPE. In \cite{hwang2012ultra,li2013integrating}, radio frequency (RF)-based sensors, such as urtra-wideband (UWB) and RF Identification (RFID), are also used to track workers or loads. In~\cite{martinez2020uav}, unmanned aerial vehicle (UAV) is integrated in current construction safety planning and monitoring processes. \cite{fang2019deep, shen2021detecting,pazari2023enhancing} uses deep-learning techniques to identify unsafe behaviors of workers, helmet of workers and workers themselves. Crane, as one of the most important heavy machinery in the construction site, the research about crane safety monitoring is popular. \cite{jiang2024unsafe} introduces a recognition framework based on transfer learning to identify unsafe hoisting behaviors of tower cranes. \cite{pazari2023enhancing} presents a method that uses computer vision and deep learning algorithms to detect crane loads and workers. \cite{golcarenarenji2022machine} proposes a new machine learning empowered solution, CraneNet, to increase the visibility of a crane operator. Based on the above research, we can observe that crane safe monitoring based on artificial intelligence (AI) and vision is the current emerging research direction. 

In addition, the detection of the interested objects and behaviors is usually one of such system's core functions. In recent years, convolutional neural network (CNN) -based methods, such as region-based CNN (RCNN), Fast-RCNN, Faster-RCNN, and Mask-RCNN have been applied to object detection due to their accuracy. However, while these methods offer high precision, they often struggle with real-time performance due to their reliance on multiple stages of processing. To address these limitations, the YOLO (You Only Look Once) series was developed~\cite{terven2023comprehensive}, which simplifies the object detection process by framing it as a single regression problem, enabling faster and more efficient detection without sacrificing accuracy. This innovative approach has various applications, including the construction site safety monitoring~\cite{pazari2023enhancing}.

However, most of the safety monitoring system and corresponding object detection are based on 2D images since RGB cameras are cost-effective, easy to install and visualize, and commonly used for safety monitoring purpose~\cite{sutjaritvorakul2020simulation,chen2017real}. Even though there are depth or stereo cameras which can obtain depth information, they have limitations on effective range and precision. LiDAR is another widely used sensor for safety monitoring system which can represent object geometry in the form of 3D points~\cite{jacobsen2021real,shen2021deep}. Studies have demonstrated that point cloud data enables more precise detection of objects compared with 2D methods, as it captures the spatial and geometric details of the scene. Researchers have explored various algorithms~\cite{gomez2023efficient}, including deep learning techniques, to enhance the accuracy and efficiency of human and object detection. In recent years, it has made progress in the field of autonomous driving~\cite{qian20223d,mao20233d}, however, the application of point cloud-based 3D object detection to construction sites is still limited~\cite{chen2018performance,xu2023approach}. Such techniques are potential for safety monitoring, challenges, such as data density, computational complexity, and environmental factors, continue to impact the effectiveness of this methodology. Fusion-based 3D object detection has emerged as a powerful technique that combines information from camera and LiDAR to enhance the accuracy and robustness of object detection systems. This approach represents state-of-the-art detection solution but the application to the construction sites is also limited~\cite{shen2021deep}.

For MiC, it is a rapid and sustainable solution for construction projects but also poses some critical safety hazards on construction sites. Cranes are the most widely utilized equipment used in MiC projects but research carried out on the safety issues of cranes in such projects is limited. \cite{mohandes2022occupational} develops a crane safety index (CSI) to improve the occupational health and safety (OHS). \cite{xu2022analysis} introduces relevant machine learning methods for early warning analysis on the safety risk of the tower crane hoisting operation of prefabricated components, but this paper focuses more on theoretical analysis. An innovative tower crane path planning system for assisting crane operators in high-rise MiC is developed by \cite{zhu2022innovative}. Even though a lot of effort is working on reducing the safety risk about MiC, there is no specifically designed safety monitoring system  for MiC lifting.

\section{Method}\label{sec:method}
\begin{figure}[!htb]
    \centering
    \includegraphics[width=12 cm]{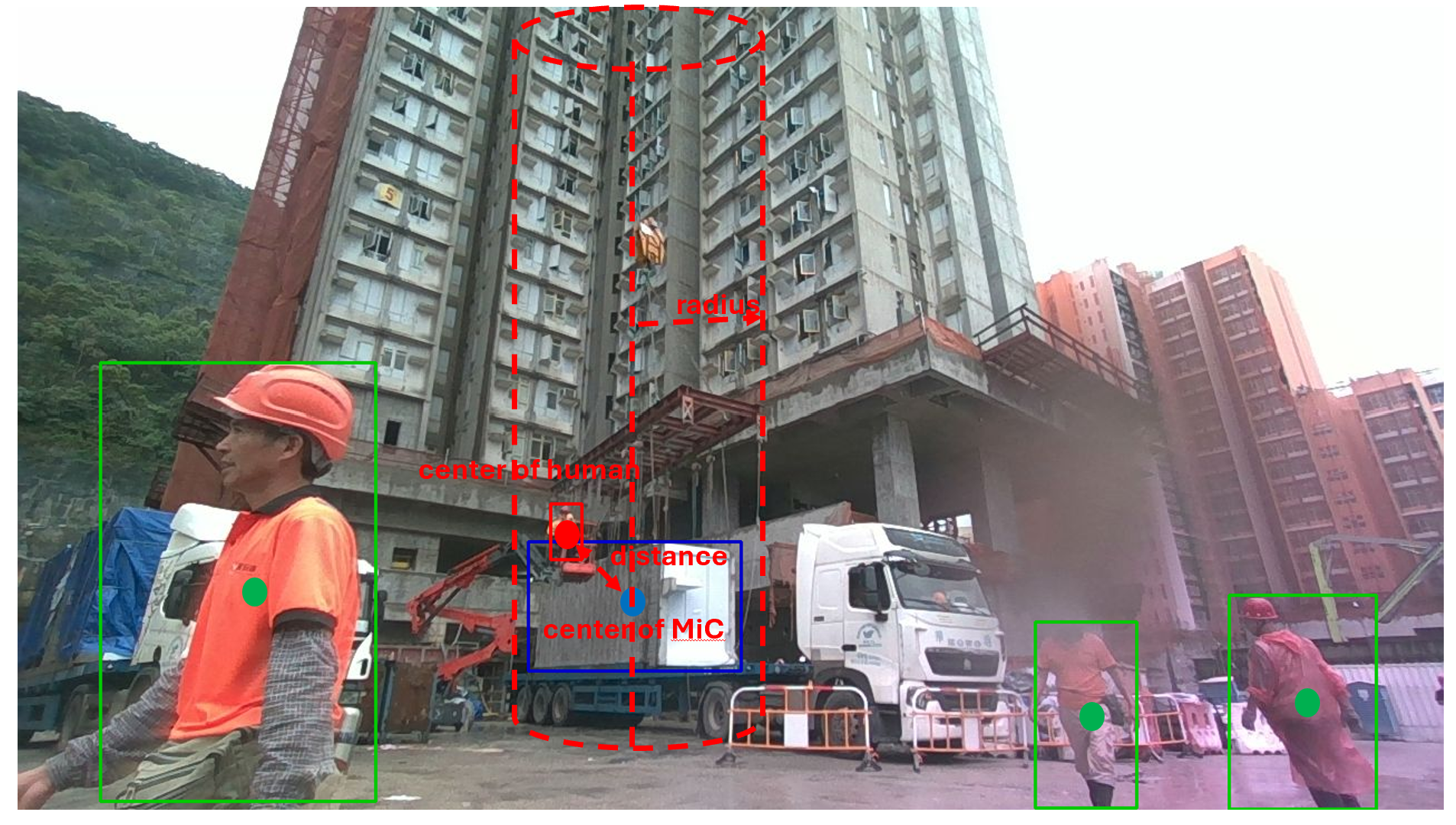}
    \caption{MiC lifting danger zone. The danger zone is a cylinder with radius $r$. The axis of the cylinder pass through the center of the MiC and is perpendicular to the ground.}
    \label{fig:MiC lifting danger zone}
\end{figure}

The Fig.~\ref{fig:pipeline} illustrates the algorithm pipeline of the developed safe lifting monitoring system that integrates data from camera and LiDAR inputs to enhance situational awareness in construction environments. Data processing and synchronization is for point cloud denoising and downsampling and time synchronization between image and point cloud. The Object detection is to identify 2D bounding boxes of MiC and humans within the scene. Target object depth, which is obtained by clustering, together with 2D bounding boxes information, will be convert to 3D coordinates in the world frame. The audible and visual alarm will be triggered if the safety risk detection block detects human in the MiC lifting danger zone as shown in Fig.~\ref{fig:MiC lifting danger zone}.  Meanwhile, the video stream with detected information will be sent and shown on cockpit GUI to provide the operator with real-time visual feedback.

\subsection{Object Detection}
Due to the advantages in detection accuracy and efficiency , we choose YOLO for 2D object detection. The baseline of the object detection model in this paper is YOLOv5~\cite{Aydin_Singha_2023}, which is introduced in 2020 by Ultralytics. Since it is convenient to train and deploy and also achieves good performance in COCO benchmark dataset, this model has been widely used in industry. The architecture of YOLOv5 mainly consists of three components: the backbone, neck and head. 

We empirically choose YOLOv8, which has better performance compared to YOLOv5, as the final object detection method in this paper. The structure of YOLOv8 consists of two components~\cite{varghese2024yolov8}: the backbone network and the detection head. The backbone network is based on EfficientNet, which is also used for extracting assortment of rich features from the input pictures at numerous scales. The head is based on NAS-FPN, which is a neural architecture search method that automatically generates feature pyramid networks for object detection. And the head is responsible for merging these features and creating different predictions of bounding boxes. Utilizing YOLOv8 for 2D object detection involves several key steps: data collection, dataset preparation, training, and inference. First, a diverse set of images depicting MiC lifting on construction sites is gathered, ensuring that the data reflects various conditions such as lighting, angles, and backgrounds to enhance model robustness. Dataset preparation includes annotation, data formatting, and splitting. After preparing the dataset, the YOLOv8 model is trained. Once trained, the model can perform inference to obtain 2D bounding boxes for MiCs and humans within the algorithm pipeline.


\subsection{Coordinate Transformation}
\begin{figure}[h]
\centering
\includegraphics[width=12 cm]{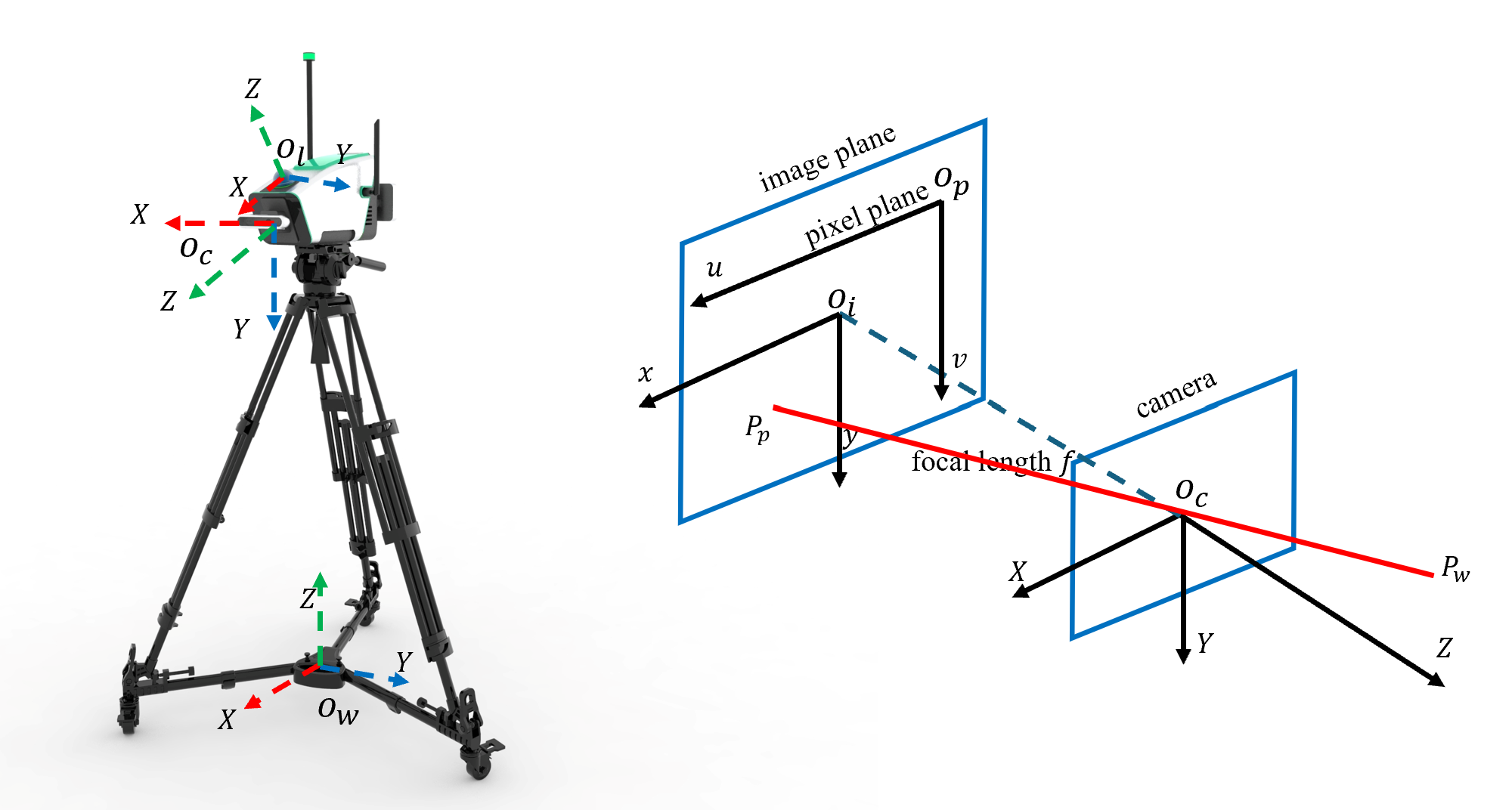}
\caption{Defined coordinate frames of the developed safe lifting monitoring system (left side) and the pinhole camera (right side). $O_w-XYZ$, $O_c-XYZ$, $O_l-XYZ$, $O_i-xyz$, $O_p-uv$ are defined as the world frame, camera frame, LiDAR frame, image frame and pixel frame respectively.}
\label{fig:frames}
\end{figure}

The Fig.~\ref{fig:frames} shows defined coordinate frames of the system and the pinhole camera model. The frames $O_w-XYZ$, $O_c-XYZ$, $O_l-XYZ$, $O_i-xyz$, $O_p-uv$ are defined as the world frame, camera frame, LiDAR frame, image frame and pixel frame respectively. We define $P_w=[x_w,y_w,z_w]^T \in \mathbb{R}^{3}$, $P_c = [x_c,y_c,z_c]^T \in \mathbb{R}^{3}$, $P_l= [x_l,y_l,z_l]^T \in \mathbb{R}^{3}$, $P_p = [u,v]^T \in \mathbb{R}^{2}$ as coordinates center of detected object represented in world frame, camera frame, LiDAR frame and pixel frame respectively. \eqref{eqn:trans_c2p} denotes the coordinate transformation between  pixel frame and camera frame and $K$ is the intrinsic matrix
\begin{equation} \label{eqn:trans_c2p}
        \left[\begin{matrix}
         P_p \\  1 \\
         \end{matrix} \right]
         = \frac{1}{Z_c} KP_c,~
        K = \left[\begin{matrix}
         f_x & 0 & c_x \\
         0   & f_y & c_y  \\
         0   & 0 & 1 \\
         \end{matrix} \right]
\end{equation} 
where $f_x$ and $f_y$ are focal lengths of the camera in the $x$ and $y$ directions while $c_x$ and $c_y$ represent the coordinates of the optical center of the camera on the image plane. The coordinate transformation between LiDAR frame and camera frame is shown in \eqref{eqn:trans_l2c}
\begin{equation} \label{eqn:trans_l2c}
        \left[\begin{matrix}
         P_c \\  1 \\
         \end{matrix} \right] 
         = T^c_l 
         \left[\begin{matrix}
         P_l \\  1 \\
         \end{matrix} \right],~ 
         T^c_l =         
         \left[\begin{matrix}
         R^c_l &  t^c_l \\ 
         0      &  1 
         \end{matrix} \right] 
\end{equation}
where $T^c_l$ is the homogeneous matrix which consists of rotation $R_l^c$ and translation $t_l^c$. The coordinate transformation between LiDAR frame and camera frame is shown in \eqref{eqn:trans_w2l} 
\begin{equation}\label{eqn:trans_w2l}
        \left[\begin{matrix}
         P_l \\  1 \\
         \end{matrix} \right] 
         = T^l_w
         \left[\begin{matrix}
         P_w\\  1 \\
         \end{matrix} \right], ~ 
         T^l_w =         
         \left[\begin{matrix}
         R^l_w &  t^l_w \\ 
         0      &  1 
         \end{matrix} \right] 
\end{equation}
where $T^l_w$ is the homogeneous matrix which consists of rotation $R^l_w$ and translation $t^l_w$. 

For depth image generation, the points from point cloud will be converted from LiDAR frame to camera frame via \eqref{eqn:trans_l2c}, and then converted to pixel frame via \eqref{eqn:trans_c2p}. For detected objects, the coordinates in pixel frame and corresponding depth are obtained from object detection and clustering algorithms. Through combining \eqref{eqn:trans_c2p}, \eqref{eqn:trans_l2c} and \eqref{eqn:trans_w2l}, such pixel coordinates and depth can be converted to coordinates in world frame through following equation
\begin{equation} \label{eqn:trans_p2w}
        \left[\begin{matrix}
         P_w \\  1 \\
         \end{matrix} \right] 
         = T^w_l T^l_c
         \left[\begin{matrix}
         Z_c K^{-1} \left[\begin{matrix}P_p \\ 1\end{matrix}\right] \\  
         1 \\
         \end{matrix} \right].
\end{equation}
Knowing the 3D coordinates of the MiC and the surrounding individuals allows us to detect if there is human in the MiC lifting danger zone as shown in Fig.~\ref{fig:MiC lifting danger zone}, and trigger alerts accordingly. 

\subsection{Clustering Algorithm}

The depth of the detected object is obtained from clustering algorithms, i.e., averaging, K-Means, and Mean-Shift, if the depth figure and bounding boxes are known. We first remove the outliers in the depth values within the bounding boxes. We will check whether there are other objects blocking the MiC. If not, we will cluster the foreground and background. If there are, we initialize three cluster centers, which represent the foreground, midground, and background depths. Next, we take the median of the foreground and midground depth values as the target depth for two cases respectively.

\section{Data processing}~\label{sec:data_processing}
In the learning-based detection task, the quality of the dataset directly determines the detection performance. On construction sites, a good dataset should have the following characteristics: 1) Size: The dataset should have a large size to train robust models in order to capture various patterns and reducing overfitting. 2) Diversity: The dataset should include a wide range of MiCs and humans in various construction sites to ensure the model can generalize well across different scenarios. 3) Variability: The dataset should include variations in conditions (e.g., lighting, angles) to make the model robust against changes in the environment. Due to the consideration of construction schedule and safety, collecting datasets that meet these conditions is challenging and is one of our core contributions.

\subsection{Calibration}
As we mentioned before, a well customized dataset is the core challenge for the object detection in our scenario. Given that the pipeline in this paper is a fusion-based method, the images from the camera and point cloud (PC) from the LiDAR are required to collected as pairs for the dataset. Two sensor units which has various performance for data collection are used: camera RealSense D455 and the LiDAR Livox MID360 , and camera Hikrobot MV-CS200-10GC and the LiDAR livox AVIA. The key parameters of the cameras and LiDARs are shown in Table.~\ref{tab: sensor units parameters}. 
\begin{table}[!htb]
    \centering
    \caption{Sensor Units Details for Data collection}
    \label{tab: sensor units parameters}
    \begin{tabular}{ccccc}
       \toprule
       Parameters & D455 + MID360 & Hikrobot + AVIA \\
       \midrule
        Camera FOV & 90 $\times$ 65°(H $\times$ V) & 90 $\times$ 90°(H $\times$ V)\\
        LiDAR FOV  & 90 $\times$ 59°(H $\times$ V) &  70.4 $\times$ 77.2° (H $\times$ V)\\
        Camera frame rate& 30 FPS& 5.9 FPS\\
        LiDAR frame rate & 10 FPS& 10 FPS\\
        Image resolution &   1280 × 800  & 5472 × 3648\\
        Detection range  &   0.1-40m     & 1- 460 m\\
       \bottomrule
    \end{tabular}
\end{table}

Two sensor units both need calibration of intrinsics and extrinsincs. We use MATLAB camera calibrator for intrinsic parameters calibration for RealSense and Hikrobot cameras. Extrinsic calibration of LiDARs and cameras is followed by the novel method in \cite{yuan2021pixel}. Through aligning edge features observed by both LiDAR and camera sensors, such approach does not require checkerboards but can achieve pixel-level accuracy. We use two sensor units with different performance for data collection in order to obtain images and point clouds with varying resolutions. However, in our system design, we chose to use the sensor unit with RealSense D455 and Livox MID360, considering the cost implications for future commercialization. The calibrated intrinsic and extrinsic matrices are given in \eqref{D455_intrinsic_extrinsics} and \eqref{Hikrobot_intrinsic_extrinsics} of Appendix section respectively.




\subsection{Data Collection}
\begin{figure}[h]
    \centering    
    \subfigure[Hong Kong-Shenzhen Innovation and Technology Park 1A construction site on Lok Ma Chau Loop]{
    \begin{minipage}[t]{6cm}
    \includegraphics[width=6cm,height=4cm]{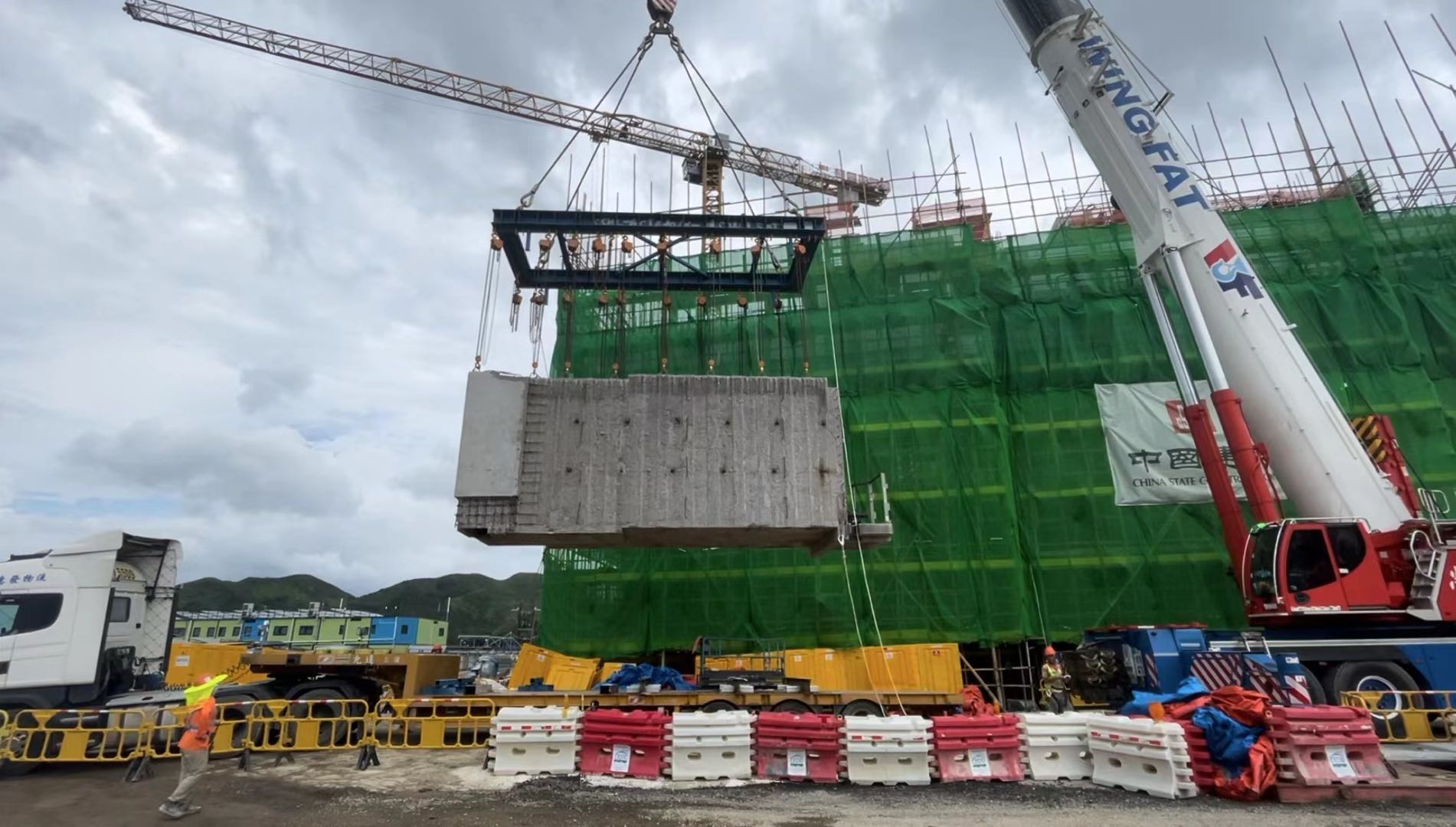}
    \end{minipage}
    }%
    \subfigure[Public Housing Developments at constructions sites
R2-6 and R2-7 on Anderson Road Quarry.]{
    \begin{minipage}[t]{6cm}
     \includegraphics[width=6cm,height=4cm]{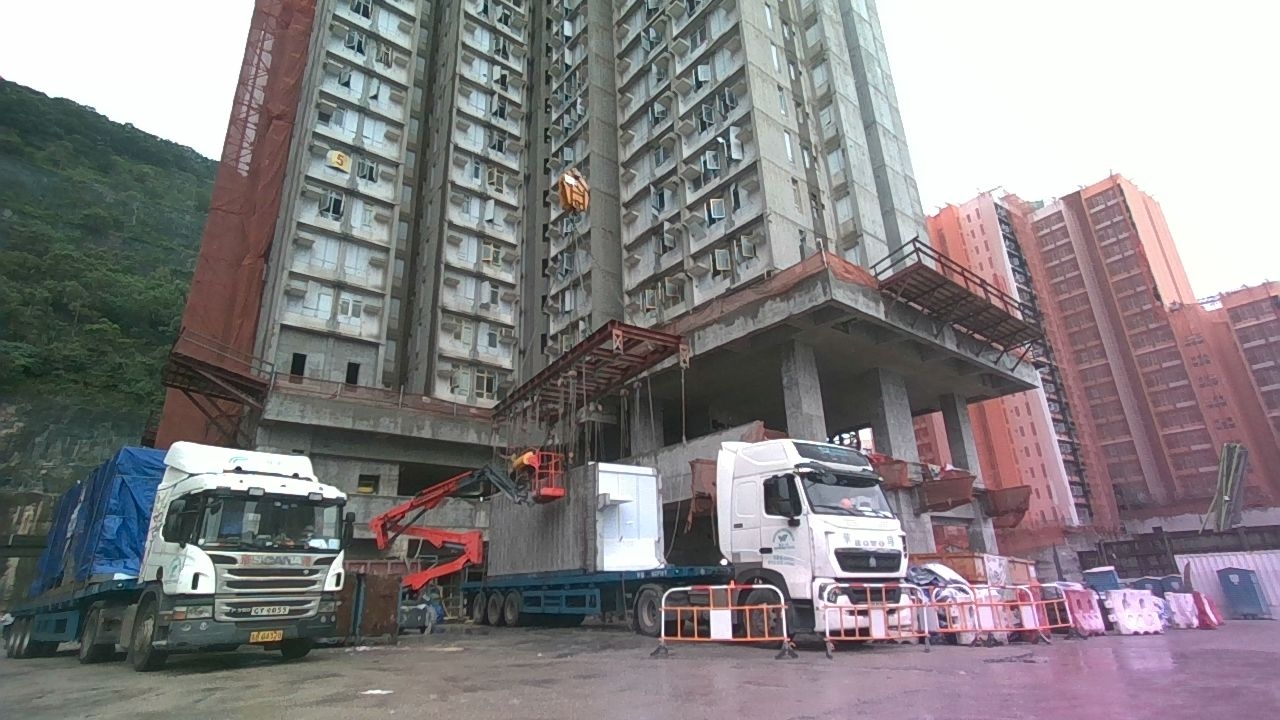}
    \end{minipage}
    }%
    \caption{Construction sites for data collection.}
    \label{fig:construction_sites}
\end{figure}

We have conducted the data collection on two construction sites in Hong Kong as shown in Fig.~\ref{fig:construction_sites}: the first one is Hong Kong-Shenzhen Innovation and Technology Park 1A on Lok Ma Chau Loop while another one is Public Housing Developments at Sites R2-6 and R2-7 on Anderson Road Quarry. The detailed data collection information is shown in Table.~\ref{tab:data_collection_info}. We select 1007 image-PC pairs in total for lifting 37 MiCs. We split the data into train (807 images), val (100 images), and test sets (100 images) and label four classes (hook, MiC, MiC frame and human) in each image. Hook and MiC frame are on the top of the MiC and used for locating the position of MiC if the detection result is not stable. 
\begin{table}[!htb]
    \centering
    \caption{Datasets Details}
    \label{tab:data_collection_info}
    \begin{tabular}{cccc}
       \toprule
       Parameters/Split & Train & Val & Test\\
       \midrule
        number of pairs & 800 & 157 & 50 \\
        number of hooks & 680 & 122 & 50\\
        number of MiCs  & 783 & 137 & 67\\
        number of MiC frames & 1083 & 233 & 55\\
        number of humans& 1831 & 392 & 111\\
       \bottomrule
    \end{tabular}
\end{table}

\begin{table*}[h]
    \centering
    \setlength\tabcolsep{1.5pt}
    \fontsize{9pt}{10pt}\selectfont
    \caption{Data Collection Details}
    \label{tab:data_collection_info}
    \scalebox{0.85}{\begin{tabular}{ccccc}
       \toprule
       Construction site & Location & Time & Sensors & MiC  \\
       &&&&lifting counts \\
       \midrule
       Hong Kong-Shenzhen Innovation and Technology Park 1A & Lok Ma Chau Loop&  Jun.12.2024 & D455+MID360  &  4 \\
       Hong Kong-Shenzhen Innovation and Technology Park 1A & Lok Ma Chau Loop &  Jun.12.2024 & Hikrobot+AVIA  &  5 \\
       Public Housing Developments at Sites R2-6 and R2-7 & Anderson Road Quarry &  Jun.14.2024 & D455+MID360 & 5 \\
       Public Housing Developments at Sites R2-6 and R2-7 & Anderson Road Quarry &  Jun.14.2024 & Hikrobot+AVIA  & 5 \\
       Public Housing Developments at Sites R2-6 and R2-7 & Anderson Road Quarry &  Aug.17.2024 & D455+MID360 & 18  \\
       \bottomrule
    \end{tabular}}\\
\end{table*}

\section{Experiment} \label{sec:experiment}
\subsection{The Experimental Platform}

\begin{figure}[h]
    \centering    
    \subfigure[Product rendering of safe lifting monitoring system.]{
    \begin{minipage}[t]{6cm}
    \includegraphics[width=6cm,height=4cm]{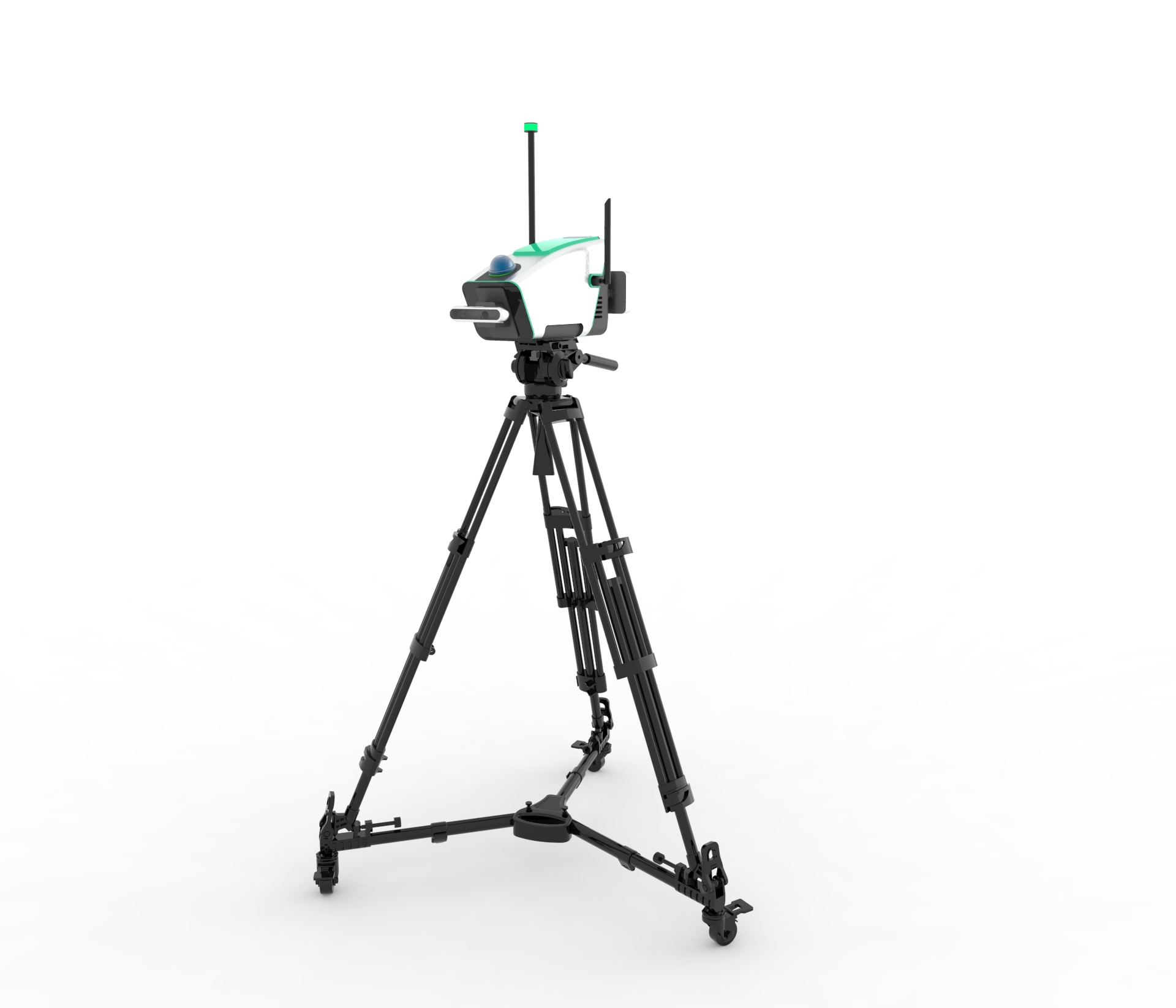}
    \end{minipage}
    }%
    \subfigure[Actual safe lifting monitoring system on the construction site.]{
    \begin{minipage}[t]{6cm}
     \includegraphics[width=6cm,height=4cm]{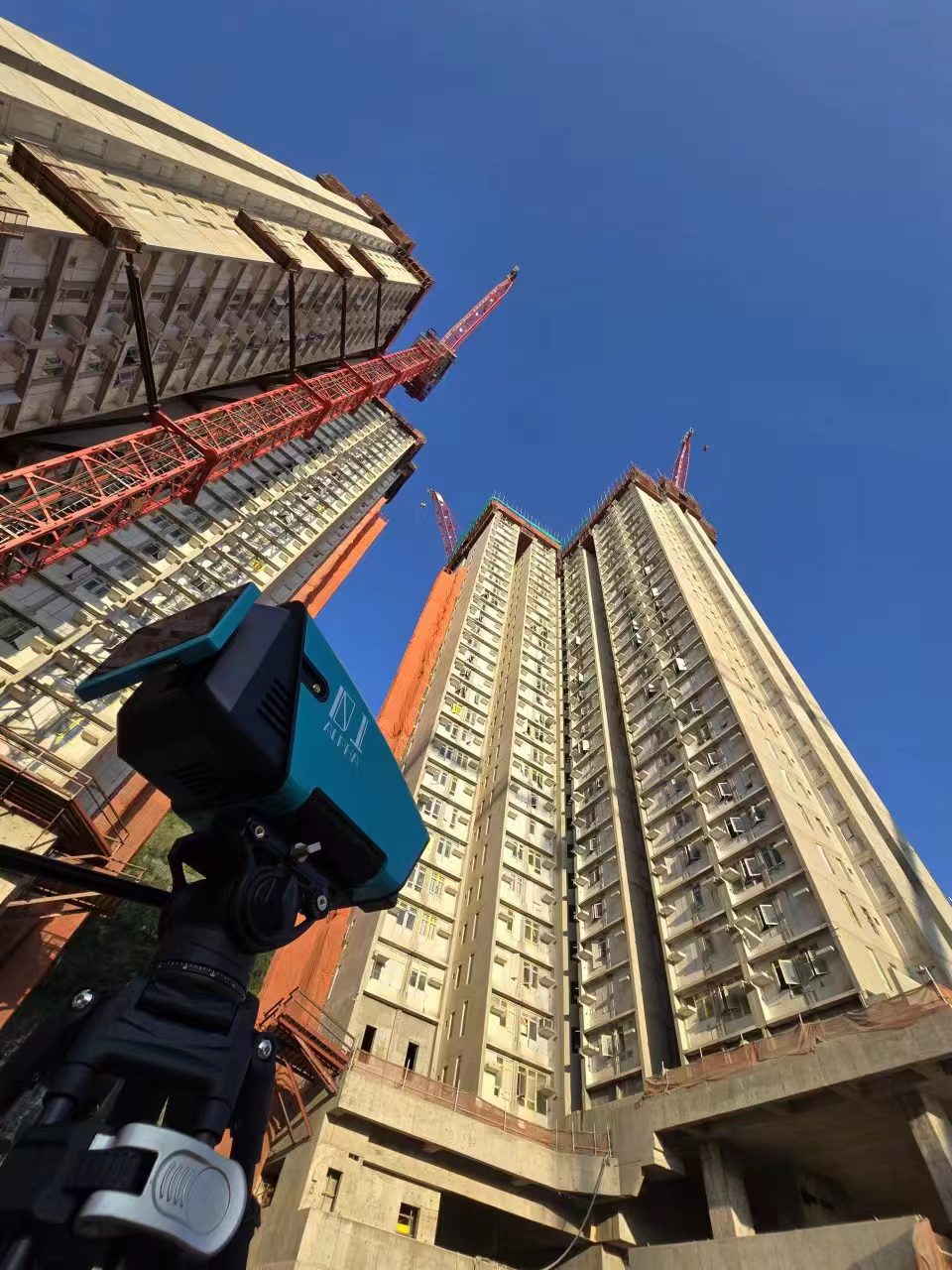}
    \end{minipage}
    }%
    \caption{The developed safe lifting monitoring system.}
    \label{fig:product}
\end{figure}
\begin{figure}[!htb]
\centering
    \includegraphics[width=10 cm]{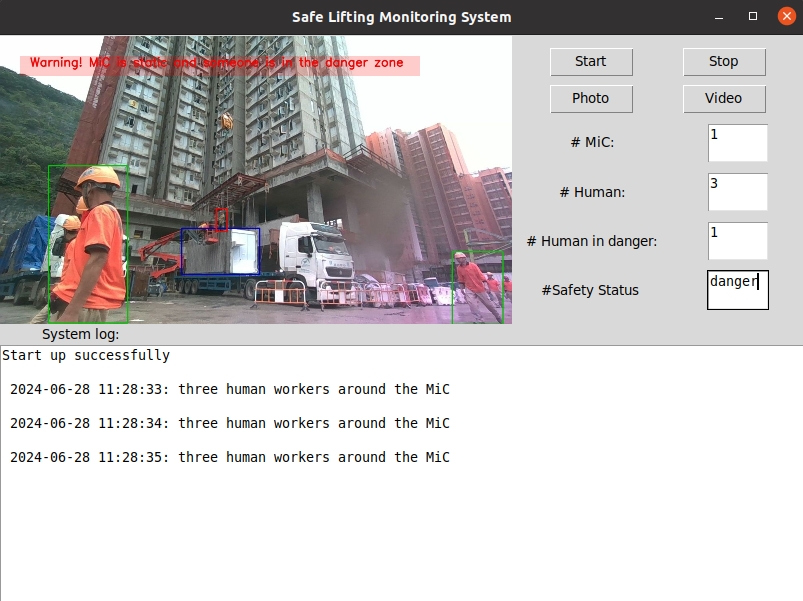}
    \caption{The GUI of the developed system.}
    \label{fig:GUI}
\end{figure}
The developed safe lifting monitoring system is shown in Fig.~\ref{fig:product}, which mainly includes a sensor unit, host processor unit, alarm unit, power supply, and graphical user interface (GUI). The sensor unit includes a RealSense D455 camera and a Livox MID360 LiDAR. Since the deep learning inference requires the strong computing ability, we choose NVIDIA Jetson Orin NX as the host processor unit. The alarm unit includes a flashing light and a speaker which are used for alarming the human workers if there is a safety hazard. 

The GUI, as shown in Fig.~\ref{fig:GUI}, is designed to provide an intuitive and user-friendly experience for tower crane operators. The interface prominently displays real-time camera feeds alongside safety information, highlighting detected lifting objects and workers. This comprehensive layout not only enhances situational awareness but also supports proactive safety management by allowing tower crane operator to quickly respond to identified hazards and monitor site conditions effectively.

\begin{figure}[h]
    \centering    
    \subfigure[Dangerous case]{
    \begin{minipage}[t]{6cm}
    \includegraphics[width=6cm,height=4cm]{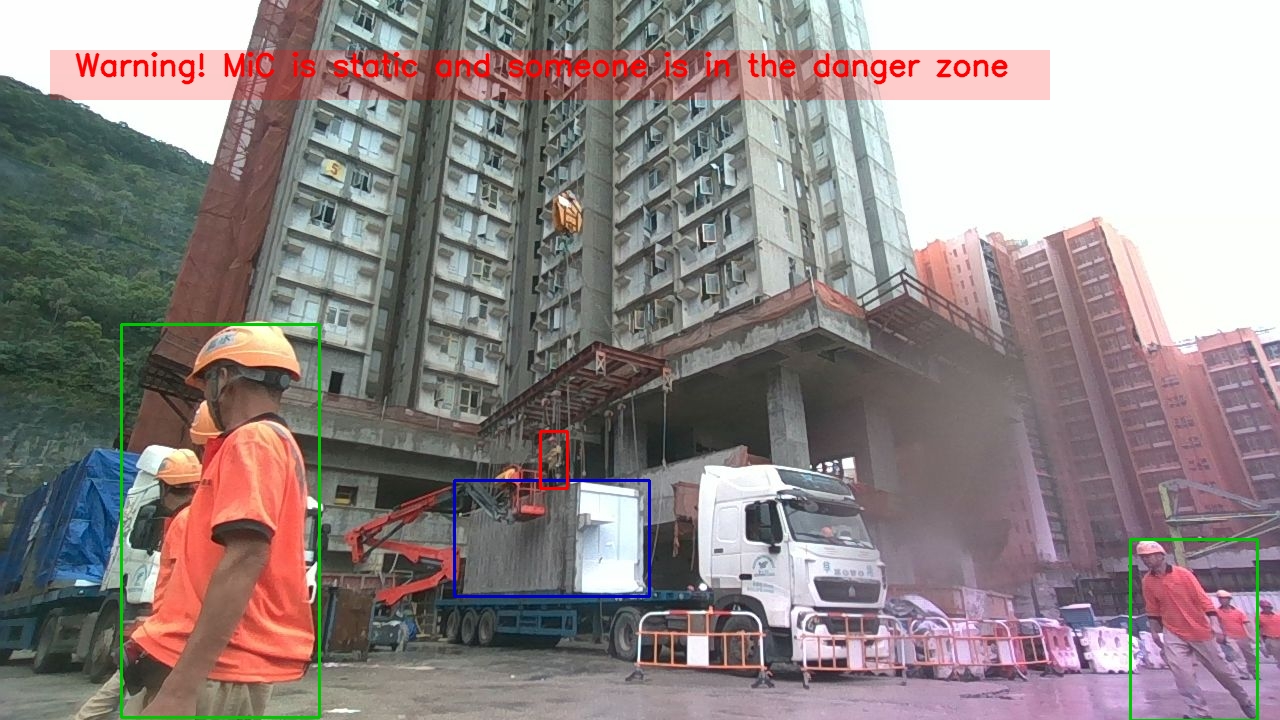}
    \end{minipage}
    }%
    \subfigure[Safe case]{
    \begin{minipage}[t]{6cm}
     \includegraphics[width=6cm,height=4cm]{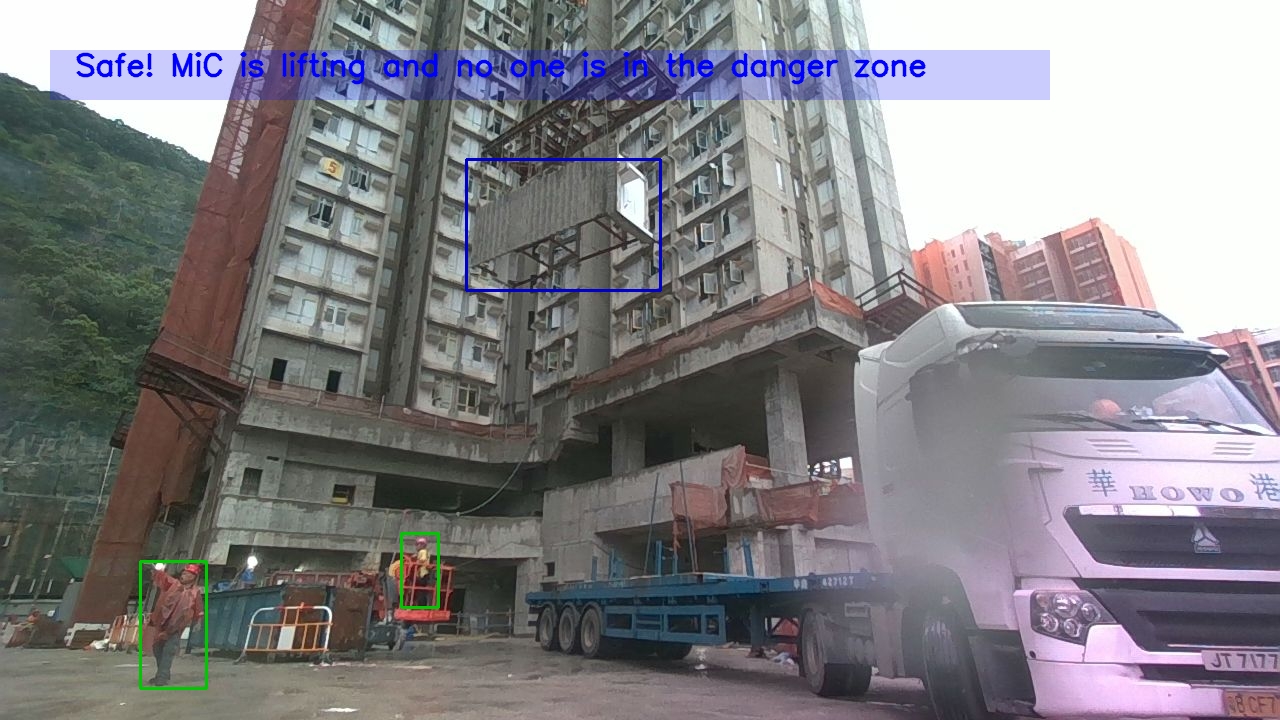}
    \end{minipage}
    }%
    \caption{Visualization of safe and dangerous cases}
    \label{visualization of safe and danger cases}
\end{figure}

\begin{small}
\begin{table*}[!htb]
    \centering
    \caption{Performance comparison with advanced algorithms}
    \label{tab:YOLOv5_vs_YOLOv8}
    \scalebox{0.85}{\begin{tabular}{ccccccc}
       \toprule
       Parameter $\&$ metrics (unit)  & YOLOv5s & YOLOv5m & YOLOv5l & YOLOv8s & YOLOv8m & YOLOv8l\\
       / Model & & & & & & \\
       \midrule
       Image size(pixels) & 640 & 640 & 640 & 640 & 640 & 640\\
       Batch size & 16 & 16 & 16 & 16 & 16 & 16 \\
       Epoch      & 300 & 300 & 300 & 300 & 300 & 300 \\
       Parameters (M)  & 7.2 & 21.2 & 46.5 & 11.2 & 25.9 & 43.7\\
       mAP50      & 0.558 & 0.538  & 0.58 &  0.852 & 0.857 & 0.864\\
       mAP50-95   & 0.243 & 0.227  & 0.239 & 0.468 & 0.485 & 0.5\\
       \bottomrule
    \end{tabular}}\\
\end{table*}
\end{small}

\subsection{The Analysis and Results}


In classification, the terms true positive (TP), false positive (FP), and false negative (FN) are commonly used to describe samples that are correctly identified, incorrectly identified, and not identified, respectively. This paper selects some commonly used metrics, namely precision (P), recall (R), average precision (AP), mean average precision (mAP),  to evaluate performance of trained models. The equations for evaluation metrics are shown below
\begin{equation}
    P = \frac{TP}{TP + FP},
\end{equation}
\begin{equation}
    R = \frac{TP}{TP + FN},
\end{equation}
\begin{equation}
    AP = \int_{0}^{1}P(r)dr,
\end{equation}
\begin{equation}
    mAP = \frac{\sum_{i=1}^{N}AP_i}{N},
\end{equation}
where $N$ is the total number of classes. The input image size of the model is adjusted to $640 \times 640$, with the batch size set at 16 and the number of iterations set at 300 epochs. 

\begin{table}[!htb]
    \centering
    \caption{YOLOv8s object detection performance}
    \label{tab:YOLOv8s performance}
    \begin{tabular}{ccccc}
       \toprule
       Class & Precision & Recall   & mAP50 & mAP50-95\\
       \midrule
        All       & 0.82 & 0.87 & 0.85 & 0.49 \\
        Hook      & 0.96 & 0.93 & 0.97 & 0.45 \\ 
        MiC       & 0.54 & 0.87 & 0.70 & 0.51\\
        MiC frame & 0.93 & 0.90 & 0.93 & 0.64\\
        Human    & 0.85 & 0.77 & 0.81 & 0.34\\
       \bottomrule
    \end{tabular}
\end{table}

As mentioned before, considering the accuracy and speed of the object detection, we empirically choose YOLOv8 and YOLOv5 series object detection models with small, medium and large sizes respectively.  The performance comparison of total 6 models are shown in Table.~\ref{tab:YOLOv5_vs_YOLOv8}. We can observe that the performance of YOLOv8 series models is significantly better than that of YOLOv5 series models, even though the size of YOLOv5 and YOLOv8's small, medium and large models is similar. In addition, for both YOLOv5 and YOLOv8 series models, the increase in model size does not significantly improve the effect of object detection.  Thus,we choose to use small size model of YOLOv8, YOLOv8s, in our application scenario. In Table.~\ref{tab:YOLOv8s performance}, it shows the object detection performance for YOLOv8s. The results show that the YOLOv8s model achieves an overall Precision of 0.82 and Recall of 0.87, indicating a strong ability to accurately detect and classify objects. The mAP50 and mAP50-95 metrics are 0.85 and 0.49, further demonstrating the model's strong performance. 

For the detected MiC and humans, the quality of the clustering results will affect the accuracy of the depth value, and then affect the accuracy of the three-dimensional position of the target. We select 10 frames of point clouds from the MiC lifting process, manually annotate the 3D bounding boxes for the MIC, and use the depth at the center of the 3D boxes as the ground truth. The evaluation metric of the ground truth depth and clustered depth is the root mean square error (RMSE). For K-Means, it achieves the RMSE at 1.3067 m, which is better than the RMSE at 1.7607 m of the averaging method.

\begin{table}[!htb]
    \setlength\tabcolsep{3pt}
    \centering
    \caption{Ground truth and detected 3D coordinates in the world frame of MiC and Distance error between them.}
    \label{tab:MIC_pipeline_perfomance}
    \begin{tabular}{cccc}
       \toprule
       PC & Ground truth (m)  & Detection (m)  & Distance error (m)  \\
       \midrule
        PC1 &  [19.7864,~3.1799,~2.1413] & [18.0674,~2.6608,~2.5117] & 1.8335\\
        PC2 &  [19.6967,~2.7455,~3.1367] & [18.1441,~2.5724,~2.8794] &  1.5833\\
        PC3 &  [20.1343,~2.6268,~3.1132] & [18.4829,~2.5713,~3.2282] & 1.6563\\
        PC4 &  [20.5326,~2.6201,~3.6380] & [18.9648,~2.5737,~3.8346] &  1.5808\\
        PC5 &  [20.4140,~3.0618,~5.5065] & [19.5339,~2.6322,~4.7439] & 1.2412\\
        PC6 & [20.9976,~2.5670,~6.0890] & [19.9863,~2.7802,~5.5941] & 1.1459\\
        PC7 &  [21.1009,~2.3260,7.2520] & [20.0330,~2.8168,~6.4753] & 1.4087\\
        PC8 &  [20.9805,~2.9543,~7.9688] & [19.3159,~2.8278,~7.1563] & 1.8566\\
        PC9 &  [20.2446,~2.0257,8.5612] & [18.1342,~1.7088,~7.9866] &  2.2101\\
        PC10 &  [20.3859,~2.9921,~9.8540] & [19.5146,~3.2900,~9.2099] & 1.1237\\
       \bottomrule
    \end{tabular}
\end{table}

\begin{table}[!htb]
\setlength\tabcolsep{3pt}
    \centering
    \caption{Ground truth and detected 3D coordinates in the world frame of humans and distance error between them. Note that it is possible that more than one human are detected in one frame of point cloud. }
    \label{tab:Human_pipeline_perfomance}
    \begin{tabular}{cccc}
       \toprule
        PC  & Ground truth (m) &  Detection (m) &  Distance error (m)  \\
       \midrule
        PC1 &  [15.6160,~4.5763,~3.6900] & [15.4050,~4.6032,~3.6958] & 0.2128\\
         &  [11.6337,-4.7351,~0.2594] & [10.7170,-5.0400,~0.4026] & 0.9766\\
        PC2 & [15.5436,~4.5564,~3.5129] & [15.5708,~4.6468,~3.5755] &  0.1133\\
             & [11.3137,~-5.7791,~0.1208] & [11.6101,-6.4863,~0.3419] & 0.7980 \\
        PC3 &  [15.6090,~4.6974,~3.0662] & [15.2507,~4.6217,~3.2059] & 0.3920\\
             & [12.4591,-5.0864,-0.1620] & [12.2062,-6.7335,~0.3682] &  1.7487\\
        PC4 & [15.5877,~4.6382,~2.8447] & [15.3821,~4.6679,~2.8878] &  0.2122\\
             & [11.8781,-5.2704,~0.1244] & [11.2280,-5.6974,~0.3583] &  0.8122\\
        PC5 & [15.4838,4.5740,2.3712] & [14.8790,~4.5186,~2.4817] & 0.6173\\
            & [10.1014,-3.9966,~0.3729] & [10.4956,-4.2575,~0.3905] & 0.4370\\
        PC6 & [15.5885,~4.6946,~2.2795] & [15.2022,~4.6630,~2.4077] & 0.4082\\
             &  [9.7081,-2.2212,~0.4437] & [11.1605,-2.7821,~0.2784] & 1.5675\\
        PC7 &  [15.4565,4.6055,2.2937] &[15.0793,~4.5326,~2.3192]  & 0.3850\\
             &[9.7479,-0.8041,~0.5049] & [9.5776,-0.9492,~0.3560] & 0.2688\\
        PC8 & [10.4902,-0.3822,~1.0195] & [11.2542,-0.6849,~0.2696] & 1.1125\\
        PC9 & [10.5221,~0.7554,~1.0626] & [11.7527,~0.7573,~0.2649] & 1.4665 \\
        PC10 &  [15.0194,~4.4087,~1.5414] & [15.0555,~4.5402,~1.6955] & 0.2058\\
             & [11.0083,~2.3104,~0.5206] & [13.2577,~2.8281,~0.0860] & 2.3488\\
       \bottomrule
    \end{tabular}
\end{table}

To quantify the performance of the entire pipeline, we use selected 10 frames of point clouds (PCs) from the MiC lifting process as shown above. Except the MiC, We also manually annotate the 3D bounding boxes for the humans. The 3D coordinates of bounding boxes's center is regarded as the ground truth. The accuracy of the 3D coordinates of detected MiC and humans is shown in Table.~\ref{tab:MIC_pipeline_perfomance} and Table.~\ref{tab:Human_pipeline_perfomance} respectively. The mean of distance error between of MiC and humans are 1.5640 m and 0.7824 m respectively. As far as we know, The main reason of error is that the size of MIC is often quite large but the depth obtained from the point cloud corresponds to the depth of the surface rather than the depth at its center. For the distance errors related to humans, in addition to sharing the same reasons as the MIC errors, another factor is that the smaller size of humans results in fewer point clouds. Consequently, the depth estimation obtained through clustering is not sufficiently accurate.

\section{Conclusion} \label{sec:conclusion}
In this paper, we consider cranes' safe lifting monitoring problem. We develop a 3D object detection pipeline which fuse the information of the camera and LiDAR. We first use the image-based object detection algorithm to obtain bounding boxes of people and lifting objects. Meanwhile the point cloud from the LiDAR is transformed as the depth image. Though clustering of depth values in the bounding boxes, the depths of people and lifting objects are known. We collect data at 2 construction sites in Hong Kong in total. We split those data into train, val, and test sets and train multiple object detection models. The performance of various models are compared and YOLOv8s is selected in the pipeline with the consideration of the model size and detection performance. The hardware for the safe lifting monitoring system has been developed, and this hardware, equipped with the designed pipeline, is being implemented on construction sites. The effectiveness of whole system, including the hardware and software, is verified on the construction sites. The contribution of this paper mainly includes: 1) We design a three-dimensional (3D) learning-based safe monitoring pipeline for crane lifting process; 2) A dataset is created which consists of 1007 image-point cloud pairs about 37 MiC liftings along with its annotations; 3) An automated software framework is implemented and applied to our hardware system; 4) We develop and deploy such safe monitoring system and verify its effectiveness on construction sites; 5) Such system is verified to automatically work on real construction sites with minimum human participation.

The methodology shown in this paper requires the calibration of the camera and LiDAR. One of the future work will focus on the 3D object detection~\cite{qi2017pointnet} only based on the LiDAR. In addition, the current pipeline requires the process of data collection and training. Thus, the pipeline with the zero-shot, one-shot or few-shot object detection algorithms is another consideration in the future. 

\section{Acknowledgment}
We would like to thank my colleagues for their support and contributions to this research. Special thanks to Wenshan Xue for his guidance and experimental site arrangement. We also appreciate Keni Wu and Yingqi Wang for their assistance with data collection. We acknowledge the financial support from InnoHK-HKCRC and the experimental construction sites provided by the Development and Construction Division, Housing Department, Hong Kong, China and China State Construction Engineering (Hong Kong) Limited.

\appendix
\section{Appendix}
The intrinsic matrix of RealSense D455 and extrinsic matrix of RealSense D455 and Livox Mid360 are shown below:


\begin{small}
\begin{equation}\label{D455_intrinsic_extrinsics}
    \text{intrinsics}_{D455} =  \left[\begin{matrix}
         631.1799 & 0 & 641.5884 \\
         0 & 633.3630 & 362.5603 \\
         0 & 0 & 1
         \end{matrix} \right],
    \text{extrinsics}_{D455} =  \left[\begin{matrix}
         -0.0009 & -0.1000 & -0.0074 & 0.0036 \\
         0.3329 & 0.0067 & -0.9430 & -0.1063\\
         0.9430 & -0.0033 & 0.3328 & -0.0610 \\
            0   &       0 &      0 &       1 \\
         \end{matrix} \right]
\end{equation}
\end{small}
The intrinsic matrix of Hikrobot MV-CS200-10GC and extrinsic matrix of Hikrobot MV-CS200-10GC and Livox AVIA are shown below:


\begin{small}
\begin{equation}\label{Hikrobot_intrinsic_extrinsics}
    \text{intrinsics}_{Hikrobot} =  \left[\begin{matrix}
         2601.3797 & 0 & 2679.1060 \\
         0 & 2604.4639 & 1851.2654 \\
         0 & 0 & 1
         \end{matrix} \right],
    \text{extrinsics}_{Hikrobot} =  \left[\begin{matrix}
         -0.0094 & 0.0143 & -0.9999 & 0.0032\\
         -0.0030 & 0.9999 & 0.0143 &  -0.1210 \\
         0.9999 & 0.0031 & -0.0094 & 0.0338 \\
         0 & 0 & 0 & 1\\
         \end{matrix} \right]
\end{equation}
\end{small}

\begin{figure*}[!htb]
    \centering
    \includegraphics[width=16 cm]{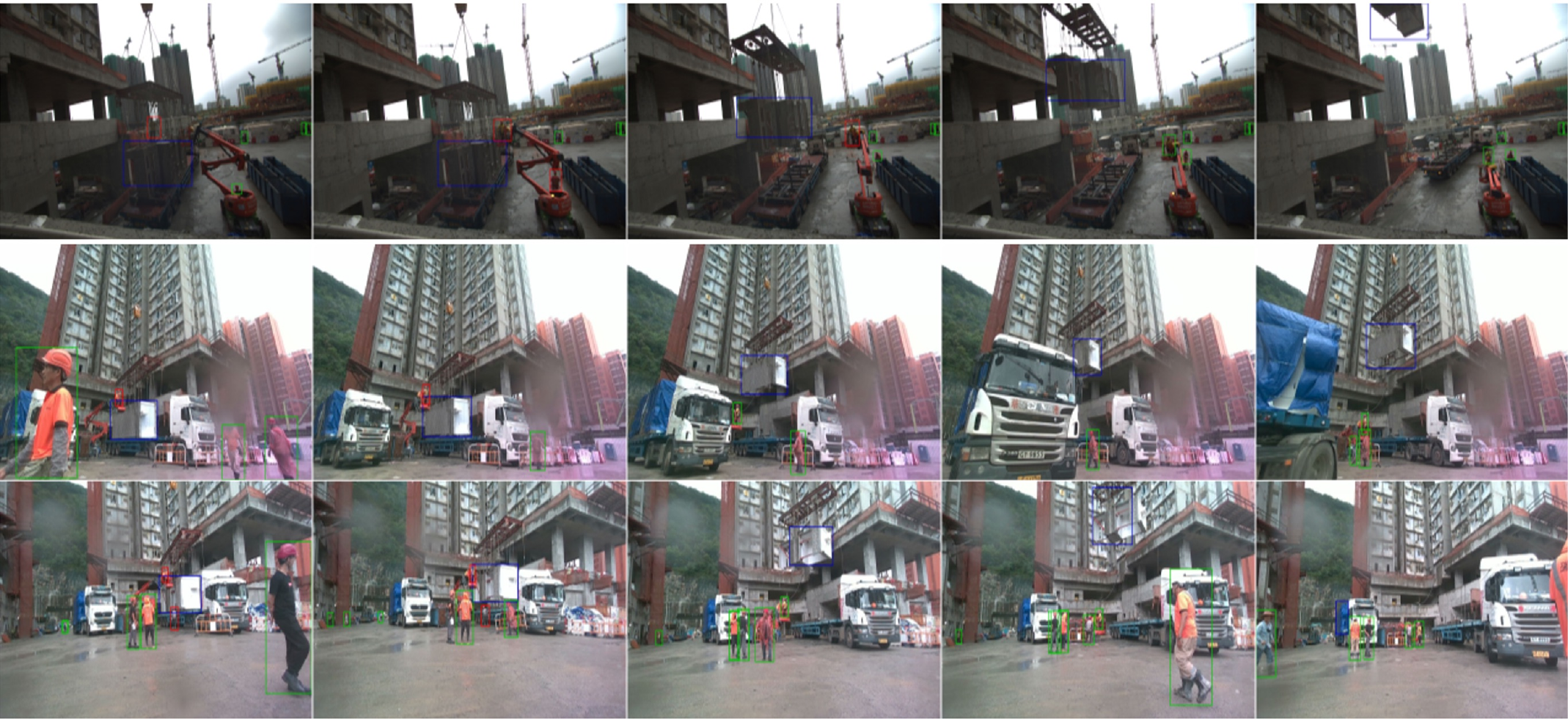}
    \caption{The MiC and humans detection results from various perspectives by YOLOv5 with IoU threshold = 0.3, Confidence = 0.5}
\end{figure*}
\begin{figure*}[!htb]
    \centering
    \includegraphics[width=16 cm]{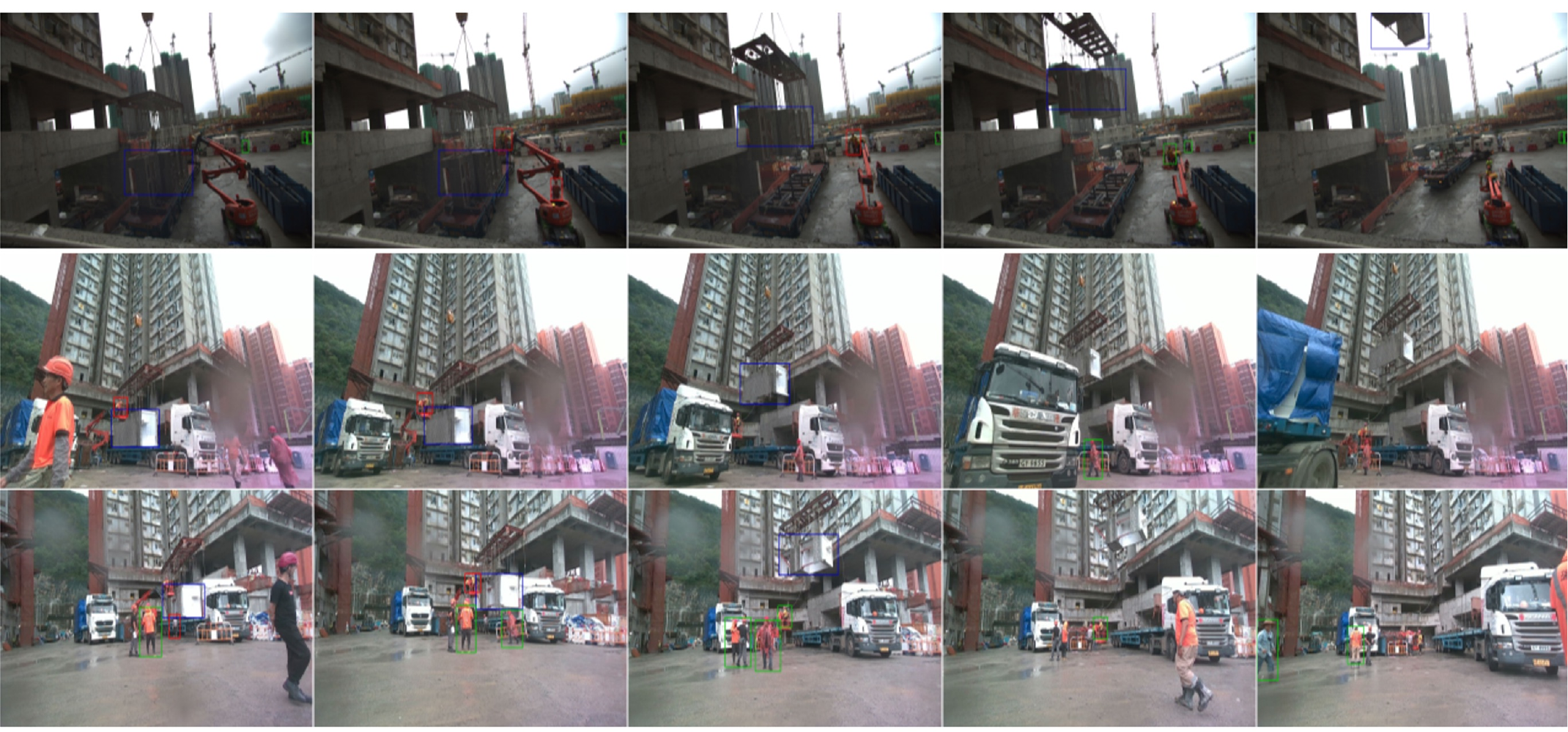}
    \caption{The MiC and humans detection results from various perspectives by YOLOv8 with IoU threshold = 0.3, Confidence = 0.5 }
\end{figure*}

\bibliography{sn-bibliography}

\end{document}